\documentclass[aps,prb,twocolumn,showpacs,preprintnumbers,amsmath,amssymb,superscriptaddress]{revtex4}

\usepackage{graphicx}
\usepackage{dcolumn}
\usepackage{bm}
\usepackage{color}

\newcommand{\bra}[1]{\langle #1|}
\newcommand{\ket}[1]{|#1\rangle}

\begin{document}

\title{Theory of shot noise in single-walled metallic carbon nanotubes
weakly coupled to nonmagnetic and ferromagnetic leads}

\author{I. Weymann}
\email{weymann@amu.edu.pl} \affiliation{Department of Physics,
Adam Mickiewicz University, 61-614 Pozna\'n, Poland}

\author{J. Barna\'s}
\affiliation{Department of Physics, Adam Mickiewicz University,
61-614 Pozna\'n, Poland} \affiliation{Institute of Molecular
Physics, Polish Academy of Sciences, 60-179 Pozna\'n, Poland}

\author{S. Krompiewski}
\affiliation{Institute of Molecular Physics, Polish Academy of
Sciences, 60-179 Pozna\'n, Poland}

\date{\today}

\begin{abstract}
We present theoretical study of shot noise in single wall metallic
carbon nanotubes weakly coupled to either nonmagnetic or
ferromagnetic leads. Using the real-time diagrammatic technique,
we calculate the current, Fano factor and tunnel magnetoresistance
in the sequential tunneling regime. It is shown that the
differential conductance displays characteristic four-fold
periodicity, indicating single-electron charging. Such a
periodicity is also visible in tunnel magnetoresistance of the
system as well as in the Fano factor. The present studies
elucidate the impact of ferromagnetic (vs. nonmagnetic) contacts
on the transport characteristics under consideration.
\end{abstract}


\maketitle

\section{Introduction}

In connection with more and more demanding miniaturization
requirements of the modern electronics, there has recently been
much interest in electric transport in low-dimensional
nanostructured materials like quantum dots and quantum wires (both
conventional \cite{kouwenhoven01,harrison99} and magnetic
\cite{wolf01,loss02,maekawa02,zutic04,maekawa06,sanvito06}).
Similarly, carbon nanotubes (CNTs) have been attracting wide
attention for more than one and a half decade due to their
exceptional transport and mechanical properties.\cite{saito98} In
particular, depending on how a graphite monolayer (graphene) is
rolled up into a cylinder, the resulting nanotube may be either
metallic or semiconducting. CNTs may be used as interconnects as
well as switching and nonvolatile memory devices in
nanoelectronics. \cite{anantram06,hoenlein} Furthermore, there is
no doubt that CNTs are also quite promising for spintronic
applications, because when placed between ferromagnetic electrodes
they show a considerable giant (tunnel) magnetoresistance, GMR
(TMR), ranging from a few, up to several tens percent.
\cite{tsukagoshi99,zhao02,kimPRB02,sahoo05,jensenPRB05,manPRB06,
liuPRB06,nagabhiravaAPL06,krompiewski05,krompiewski06,
cottetPRB06,cottet06,schonenberger06} Moreover, recent experiments
on single wall CNTs with moderate coupling to external leads
\cite{sahoo05} have shown that TMR in nanotubes can become
negative. This behavior was accounted for theoretically in terms
of spin-dependent phase shifts at interfaces. \cite{cottetPRB06}
Therefore our theoretical analysis of CNTs is extended so as to
include, on the one hand detailed intrinsic energy-level structure
of the given CNT (singlet vs. triplet energy level occupancies,
etc.), and on the other hand the effect of magnetic electrodes on
transport properties. It is now well-known that at low temperature
the Coulomb blockade regime may manifest itself even when the
so-called addition energy is less than energy-level spacings. As
shown in Refs. [\onlinecite{liangPRL02,sapmazPRB05}], the
conductance spectra of CNTs in that case reveal a characteristic
four-fold structure with relatively high peaks (of the order of
$e^2/h$).

In this paper we analyze theoretically the charge and spin
transport in single wall CNTs weakly coupled to metallic leads --
either nonmagnetic or ferromagnetic. We consider Coulomb blockade
transport regime, where the mean-field model of Oreg et al.
\cite{oregPRL00} is applicable. That is the case when the charging
energy-to-mean level spacing ratio is small enough (and
consequently the Luttinger parameter $g$ in not drastically less
than 1) - otherwise, the Luttinger-liquid model would be more
appropriate. \cite{egger,kane,bockrath} Our considerations are
based on the real-time diagrammatic technique. Assuming realistic
parameters of the system, \cite{liangPRL02,sapmazPRB05} we
calculate the basic transport characteristics, i.e. the current,
conductance, and TMR in the case of ferromagnetic contacts. Apart
from this, we also calculate the shot noise (associated with
discrete nature of charge) and show that the corresponding Fano
factor $F$ is suppressed below the Schottky value, $F<1$, beyond
the Coulomb blockade regime. More specifically, in the transport
regions where sequential contribution to the current is dominant,
the shot noise is sub-Poissonian with $F$ slightly above 0.5.
Since our considerations are limited to the first-order transport
with respect to the coupling parameter, the above conclusion does
not apply to regions where sequential tunneling processes are
exponentially suppressed. It is noteworthy that the linear and
nonlinear transport across carbon nanotubes has been recently
addressed by Mayrhofer and Grifoni.
\cite{mayrhoferPRB06,mayrhoferEPJB07} These considerations however
deal with the case of (i) nonmagnetic leads, and (ii) basically
large diameter nanotubes, for which both the exchange effects and
back scattering processes can be neglected. Spin-polarized
transport for collinear and noncollinear alignments of the leads'
magnetic moments was discussed in Refs.
[\onlinecite{cottet06,koller07}]. However, the considerations
presented in Ref. [\onlinecite{koller07}] are again subjected to
the constraints (ii), whereas Ref. [\onlinecite{cottet06}]
concerns only the linear response regime (zero bias limit). In the
present paper we focus mainly on the shot noise, using a
nonequilibrium approach applicable, in principle, to carbon
nanotubes of arbitrary diameter. It seems that, at present,
theoretical studies on the shot noise in CNTs lag behind recent
experimental ones (e.g. [\onlinecite{onac06, wu07, tsuneta07}]).
On the other hand, we are not aware of any investigations (either
theoretical or experimental) on the shot noise in
ferromagnetically contacted CNTs. Therefore, the main objective of
this study is to fill, at least partially, these gaps. In
addition, we also believe that our considerations will stimulate
experimental research on CNTs weakly coupled to ferromagnetic
leads, and will be helpful in understanding future experiments.

The paper is organized as follows: In Sec. II the model and
computational method are shortly described. Sections III and IV
present results on electronic transport (including the shot noise)
in the sequential tunneling approximation for nonmagnetic and
ferromagnetic contacts, respectively. Finally, Sec. V provides a
short summary and conclusions.

\section{Model and method}

The system considered in this paper consists of a single wall
metallic CNT and two electron reservoirs (electrodes) which are
weakly coupled to the CNT. The electrodes can be either
nonmagnetic or ferromagnetic. In the latter case, the net spin
moments (and magnetizations) of the leads are assumed to be
collinear, i.e., they can form either parallel or antiparallel
magnetic configuration. The Hamiltonian $\hat{H}$ of the system
takes the general form
\begin{equation}\label{Eq:H}
   \hat{H}=\hat{H}_{\rm L} + \hat{H}_{\rm R} + \hat{H}_{\rm CNT} +
   \hat{H}_{\rm T} \,.
\end{equation}
The first two terms describe noninteracting itinerant electrons in
the leads,
\begin{equation}
   \hat{H}_r = \sum_{{\mathbf k}\sigma} \varepsilon_{r{\mathbf
   k}\sigma} c^{\dagger}_{r{\mathbf k}\sigma} c_{r{\mathbf
   k}\sigma}
\end{equation}
for the left ($r={\rm L}$) and right ($r={\rm R}$) lead, with
$\varepsilon_{r{\mathbf k}\sigma}$ being the energy of an electron
with the wave vector ${\mathbf k}$ and spin $\sigma$ in the lead
$r$, and $c^{\dagger}_{r{\mathbf k}\sigma}$ ($c_{r{\mathbf
k}\sigma}$) denoting the respective creation (annihilation)
operator.

The third term, $\hat{H}_{\rm CNT}$, in the Hamiltonian,
Eq.~(\ref{Eq:H}), describes the CNT and is assumed in the form
introduced by Oreg et al. \cite{oregPRL00}
\begin{eqnarray}\label{Eq:HNT}
   \hat{H}_{\rm CNT} &=& \sum_{\mu j\sigma}
   \varepsilon_{\mu j} n_{\mu j\sigma} + \frac{U}{2}
   \left[ \sum_{\mu j\sigma} n_{\mu j\sigma} - N_0 \right]^2
   \nonumber\\
   &+& \delta U \sum_{\mu j} n_{\mu j\uparrow} n_{\mu j\downarrow}
   + J \sum_{\mu j, \mu^\prime j^\prime}
   n_{\mu j\uparrow} n_{\mu^\prime j^\prime\downarrow}
   \,,
\end{eqnarray}
where $n_{\mu j\sigma} = d^{\dagger}_{\mu j\sigma}d_{\mu
j\sigma}$, and $d^{\dagger}_{\mu j\sigma}$ ($d_{\mu j\sigma}$) is
the creation (annihilation) operator of an electron with spin
$\sigma$ on the $j$th level in the subband $\mu$ ($\mu=1,2$). The
discrete structure of the energy levels results from the
quantization of electron motion along the CNT, which is an
appropriate starting point in the limit of weak coupling between
the nanotube and the leads. The corresponding energy
$\varepsilon_{\mu j}$ of the $j$th discrete level in the subband
$\mu$ is given by $\varepsilon_{\mu j} = j\Delta + (\mu-1)\delta$,
where $\Delta$ is the mean level spacing and $\delta$ describes
the energy mismatch between the discrete levels corresponding to
the two subbands. The second term in Eq. (\ref{Eq:HNT}) stands for
the electrostatic energy of a charged CNT, with $U$ denoting the
charging energy and $N_0$ being the charge on the nanotube induced
by gate voltages. The next term corresponds to the on-level
Coulomb interaction with $\delta U$ being the relevant on-site
Coulomb parameter. Finally, the last term in Eq. (\ref{Eq:HNT})
describes the exchange energy, with $J$ being the relevant
exchange parameter. Furthermore, the system is symmetrically
biased and we assume equal capacitive couplings to the left and
right leads, so the dependence of the nanotube energy levels on
the bias voltage may be neglected. The exchange effects described
by $J$ play an important role for small diameter nanotubes (up to
2 nm or so), whereas for nanotubes of larger diameters these
effects become negligible. Thus in the latter case transport
properties can be qualitatively described without taking into
account the exchange interaction. \cite{mayrhoferPRB06,
mayrhoferEPJB07} Incidentally we show in the following that $J$
happens to be rather small in the present model, too. We also
notice that the mean-field model of Oreg et al. \cite{oregPRL00},
Eq. (3), can be used to analyze the ground state properties and
particle-hole excitations of single wall carbon nanotubes, however
it fails to describe collective excitations such as spin and
charge density waves, spin-charge separation, etc. (Luttinger
liquid behavior). \cite{egger,kane,bockrath}

The last term of Hamiltonian, $\hat{H}_{\rm T}$, takes into
account tunneling processes between the CNT and electrodes, and is
assumed in the form,
\begin{equation}
  \hat{H}_{\rm T}=\sum_{r=\rm
  L,R}\sum_{\mathbf k} \sum_{\mu j\sigma}\left(t_{rj}c^{\dagger}_{r {\mathbf k}\sigma}
  d_{\mu j\sigma}+ t_{rj}^\star d^\dagger_{\mu j\sigma} c_{r {\mathbf k}\sigma}
  \right) \,,
\end{equation}
where $t_{rj}$ denotes the tunnel matrix elements between the lead
$r$ and the $j$th level (assumed to be spin-independent also for
ferromagnetic leads). The $j$th level coupling to external leads
can be described by $\Gamma_{rj}^{\sigma}= 2\pi |t_{rj}|^2
\rho_r^\sigma$, with $\rho_r^\sigma$ being the density of states
in the lead $r$ for spin $\sigma$. The role of ferromagnetic leads
is taken into account just {\it via} the spin-dependent density of
states $\rho_r^\sigma$. By introducing the spin polarization $p_r$
of the lead $r$, $p_{r}=(\rho_{r}^{+}- \rho_{r}^{-})/
(\rho_{r}^{+}+ \rho_{r}^{-})$, the coupling parameters
$\Gamma_{rj}^{\sigma}$ can be expressed as
$\Gamma_{rj}^{+(-)}=\Gamma_{rj}(1\pm p_{r})$, with $\Gamma_{rj}=
(\Gamma_{rj}^{+} +\Gamma_{rj}^{-})/2$. Here, $\Gamma_{rj}^{+}$ and
$\Gamma_{rj}^{-}$ describe the coupling of the $j$th level to the
spin-majority and spin-minority electron bands of the $r$th lead,
respectively. When the leads are nonmagnetic, then
$\Gamma_{rj}^{+}=\Gamma_{rj}^{-}$. In the following we assume
$\Gamma_{rj}=\Gamma/2$ for all values of the indices $j$ and $r$.

We analyze electronic transport through the single wall metallic
CNT described by the Hamiltonian (\ref{Eq:HNT}), and the
considerations are limited to the sequential (first order in
tunneling processes) tunneling regime. The first-order tunneling
gives the dominant contribution to the charge and spin currents
for voltages above a certain threshold voltage, and is
exponentially suppressed in the Coulomb blockade regime. The
effects due to higher-order tunneling processes, e.g. cotunneling,
\cite{cotunneling,weymannPRB07,weymannPRB05} are not taken into
account. When the bias voltage exceeds the threshold voltage, the
current flows due to tunneling of electrons one by one through the
discrete energy levels of the nanotube. In order to make the
analysis most realistic, in numerical calculations we have taken
into account up to six different orbital levels, three in each
subband, which results in $4^6$ many-body nanotube states
$\ket{\chi}$.

Transport is calculated with the aid of the real-time diagrammatic
technique. \cite{diagrams,thielmann,weymannPRB05} This technique
consists in a systematic perturbation expansion of the reduced
density matrix and current operator in the couplings
$\Gamma_{rj}$. To determine the stationary occupation
probabilities, charge current, and shot noise in the sequential
tunneling regime, we employ the matrix approach developed in Ref.
[\onlinecite{thielmann}], and introduce the respective self-energy
matrices: $\mathbf{W}$, $\mathbf{W^I}$, $\mathbf{W^{II}}$. The
matrix $\mathbf{W}$ contains self-energies with one arbitrary row
$\chi_0$ replaced by $(\Gamma,\dots,\Gamma)$, which is due to the
normalization of the probabilities, $\sum_{\chi}P_{\chi}=1$. The
elements $W_{\chi \chi^\prime}$ of the matrix $\mathbf{W}$
describe the first-order tunneling transitions between the
$\ket{\chi}$ and $\ket{\chi^\prime}$ many-body states. They are
given by, \cite{thielmann} $W_{\chi \chi^\prime}=W_{\chi
\chi^\prime} ^{\rm L} + W_{\chi \chi^\prime}^{\rm R}$, where
\begin{eqnarray*}
   W_{\chi \chi^\prime}^r &=& 2\pi \sum_{\sigma} \rho_r^\sigma
   \left\{
   f_r(\varepsilon_\chi - \varepsilon_{\chi^\prime})
   \left|\sum_{\mu j}
   t_{rj}^\star \bra{\chi}d_{\mu j\sigma}^\dagger
   \ket{\chi^\prime}\right|^2 \right.\\
   &+&\left. \left[ 1-f_r(\varepsilon_{\chi^\prime} - \varepsilon_{\chi}) \right]
   \left|\sum_{\mu j} t_{rj} \bra{\chi}d_{\mu j\sigma}
   \ket{\chi^\prime}\right|^2 \right\}
\end{eqnarray*}
for $\chi\neq\chi^\prime$ and  $W_{\chi \chi}^r = -
\sum_{\chi^\prime\neq\chi} W_{\chi^\prime\chi}^r$, with
$f_r(\varepsilon) = 1/[e^{(\varepsilon-\mu_r)/k_{\rm B}T}+1]$ and
$\mu_r$ being the electrochemical potential of the lead $r$. The
second matrix, $\mathbf{W^I}$, denotes the full self-energy matrix
with one {\it internal} vertex, associated with the expansion of
the tunneling Hamiltonian, replaced by the current operator.
Finally, the third matrix, $\mathbf{W^{II}}$, consists of
self-energies with two internal vertices replaced by the current
operator. The current operator $\hat{I}$ is defined as
$\hat{I}=(\hat{I}_{\rm R} - \hat{I}_{\rm L})/2$, with
$\hat{I}_r=-i(e/\hbar) \sum_{{\mathbf k}\sigma} \sum_{\mu
j}\left(t_{rj}c^{\dagger}_{r {\mathbf k}\sigma} d_{\mu j\sigma}-
t_{rj}^\star d^\dagger_{\mu j\sigma} c_{r {\mathbf
k}\sigma}\right)$ being the current flowing from the nanotube into
the lead $r$. The elements of the matrices $\mathbf{W^{I}}$ and
$\mathbf{W^{II}}$ can be expressed in terms of $W_{\chi
\chi^\prime}$ as, \cite{thielmann} $W_{\chi \chi^\prime}^{\rm I} =
\left[ \Theta(N_{\chi^\prime}-N_{\chi})
-\Theta(N_{\chi}-N_{\chi^\prime}) \right] \left( W_{\chi
\chi^\prime}^{\rm R} - W_{\chi \chi^\prime}^{\rm L} \right)$ and
$W_{\chi \chi^\prime}^{\rm II} = (1-2 \delta_{\chi\chi^\prime})
W_{\chi \chi^\prime}/4$, respectively, where $N_\chi = \sum_{\mu
j\sigma} n_{\mu j\sigma}$ and $\Theta(x)$ is the step function.

Having calculated the respective matrices, the stationary
probabilities can be determined from the following equation,
\cite{thielmann}
\begin{equation}\label{Eq:master}
  \left(\mathbf{W}\mathbf{P}\right)_{\chi}=
  \Gamma\delta_{\chi\chi_0}\,,
\end{equation}
where $\mathbf{P}$ is the vector containing the occupation
probabilities. In turn, the sequential current flowing through the
CNT can be calculated from
\begin{equation}\label{Eq:current}
  I=\frac{e}{2\hbar}{\rm Tr}\{\mathbf{W^I}\mathbf{P}\} \,,
\end{equation}
with ${\rm Tr}\{\mathbf{A}\}$ denoting the trace of the matrix
$\mathbf{A}$. On the other hand, the zero-frequency shot noise,
$S=2\int_{-\infty}^0 dt\left(\langle
\hat{I}(t)\hat{I}(0)+\hat{I}(0)\hat{I}(t)\rangle-2 \langle
\hat{I}\rangle^2 \right)$, is given by
\begin{equation}\label{Eq:noise}
  S=\frac{e^2}{\hbar}{\rm Tr}\{
  \mathbf{W^{II}}\mathbf{P}
  +\mathbf{W^{I}}\mathbf{\tilde{P}}\mathbf{W^{I}}\mathbf{P}
  \} \,,
\end{equation}
where $\mathbf{\tilde{P}}$ is determined from the equation
$\mathbf{W} \mathbf{\tilde{P}}=\mathbf{Q}$, with
$Q_{\chi^\prime\chi} =\left(P_{\chi^\prime}- \delta_{\chi^\prime
\chi}\right) \left(1-\delta_{\chi^\prime \chi_0}\right)$.
\cite{thielmann}

Having found the current $I$ and the zero-frequency shot noise
$S$, one can determine the Fano factor $F$, $F=S/(2e|I|)$. The
Fano factor describes the deviation of $S$ from the Poissonian
shot noise given by $S_p=2e|I|$. Now we proceed to numerical
results based on the formalism described briefly above, and we
start with the case of nonmagnetic leads.

\section{Carbon nanotubes coupled to nonmagnetic leads}

\begin{figure}[t]
  \includegraphics[width=0.75\columnwidth]{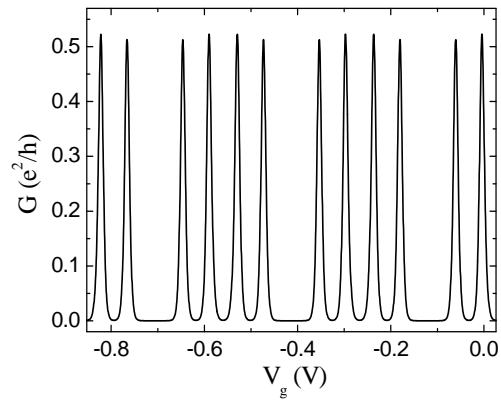}
  \caption{\label{Fig:1}
  Linear conductance as a function of the gate voltage.
  The parameters are: $\Delta = 8.4$ meV, $U/\Delta = 0.26$,
  $J/\Delta = 0.12$, $\delta U/\Delta=0.04$,
  $\delta/\Delta = 0.27$, $k_{\rm B}T/\Delta = 0.025$,
  $p_{\rm L} = p_{\rm R} = 0$, $x=0.14$,
  and $\Gamma = 0.2$ meV.}
\end{figure}

In this section we present numerical results on electronic
transport through CNTs contacted to nonmagnetic leads. The
parameters assumed to model electronic structure of the CNT-based
quantum dot weakly coupled to the leads are as follows: $\Delta =
8.4$ meV, $U/\Delta = 0.26$, $J/\Delta = 0.12$, $\delta
U/\Delta=0.04$, $\delta/\Delta = 0.27$, and $\Gamma = 0.2$ meV.
These parameters are comparable to those derived from experimental
observations. \cite{liangPRL02,sapmazPRB05} Apart from this, we
assume $k_{\rm B}T/\Delta = 0.025$, which also corresponds to the
temperature of the above mentioned experiments. In addition, to
fit our numerical data to the experimental ones presented in Refs.
[\onlinecite{liangPRL02,sapmazPRB05}], we assume equal capacitive
couplings of the CNT to the left and right leads, $C_{\rm L} =
C_{\rm R} \equiv C $, while for the gate capacitance we assume,
$C_g = xC$, with $x=0.14$.

By applying a gate voltage one can continuously shift position of
the discrete energy levels up or down, which leads to peaks in the
linear conductance each time the discrete level of the CNT crosses
the Fermi level of the leads. This can be seen in Fig.~\ref{Fig:1}
which displays the linear conductance vs the gate voltage. First
of all, one can see that the conductance spectrum exactly
reproduces the four-peak periodicity observed experimentally in
the weak coupling regime. \cite{liangPRL02,sapmazPRB05} For the
parameters assumed here the dominant energy scale is set by the
level separation $\Delta$ within individual electron subbands. The
other important energy parameters are the charging energy $U$
associated with an extra electron on the nanotube and the mismatch
between the two subbands $\delta$. The remaining parameters, i.e.
the exchange coupling $J$ and on-level Coulomb correlation $\delta
U$, are smaller and their influence on the spectra is less
pronounced, although still remarkable.

Each peak in the conductance spectrum corresponds to the addition
of a new electron to the system, which fills in a subsequent empty
energy level of the nanotube. Since there are two electron
subbands in CNTs, shifted in energy by $\delta$, and the nanotube
levels are spin degenerate, four electrons can be added to the
system before another discrete level in the same electron subband
(separated by $\Delta$) can be occupied, and a new four-peak
sequence in the linear conductance can emerge. Conductance between
the peaks is strongly suppressed, which is particularly well
pronounced for the regions between successive four-peak patterns.
The first peak in each four-peak sequence corresponds to the
situation when a new energy level becomes occupied by a single
electron. The second peak, in turn, corresponds to the case when
this level becomes occupied by another electron of opposite spin
orientation. The next two peaks appear when the energy level in
the second subband (slightly higher in energy due to a nonzero
mismatch parameter) becomes occupied first by a single electron
and then by two electrons of opposite spins. Such scenario holds
for the assumed parameters, i.e. for $J+\delta U < \delta$. When,
however, the exchange parameter becomes larger, $J+\delta U >
\delta$, the triplet state may be formed and a new scenario of the
level filing may appear. \cite{liangPRL02,sapmazPRB05}

\begin{figure}[t]
  \includegraphics[width=0.95\columnwidth]{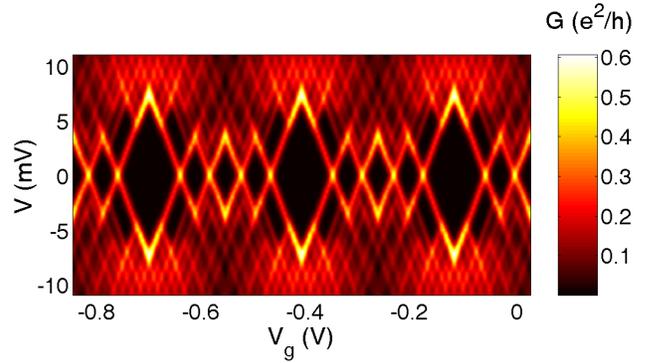}
  \caption{\label{Fig:2} (color online)
  Differential conductance as a function of bias and gate
  voltages. The parameters are the same as in Fig.~\ref{Fig:1}.}
\end{figure}

Differential conductance in both the linear and nonlinear response
regimes is shown in Fig.~\ref{Fig:2}, which clearly reveals the
blockade regions (the black diamonds) and discrete electronic
structure. Figure \ref{Fig:1} can be thus considered as a
cross-section of Fig.~\ref{Fig:2} along the line corresponding to
$V=0$. The three large diamonds in Fig.~\ref{Fig:2} correspond to
the plateaus between the successive four-peak patterns from
Fig.~\ref{Fig:1}. The smaller diamonds, in turn, correspond to the
blockade regions between the peaks inside the corresponding
four-peak sequence.

\begin{figure}[t]
  \includegraphics[width=0.65\columnwidth]{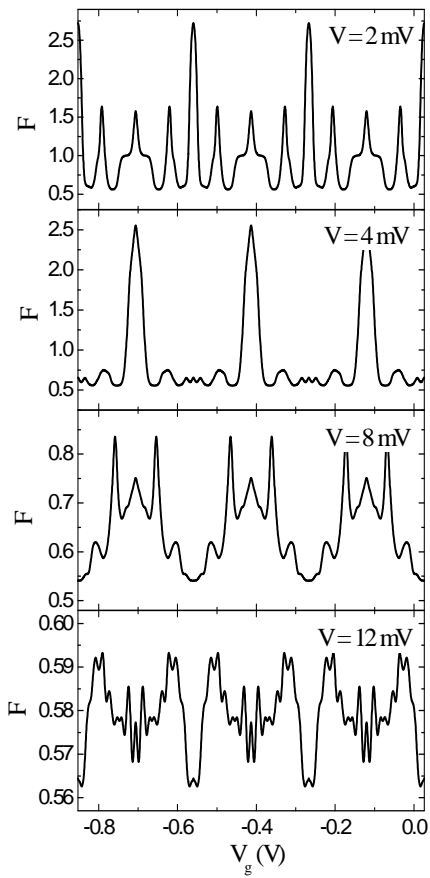}
  \caption{\label{Fig:3}
  Fano factor as a function of gate voltage
  for several values of bias voltage $V$.
  The parameters are the same as in Fig.~\ref{Fig:1}.}
\end{figure}

The Fano factor, calculated as a function of the gate voltage and
for the same parameters as in Fig.~\ref{Fig:1}, is shown in
Fig.~\ref{Fig:3} for several values of the bias voltage, $|eV|
> k_{\rm B}T$. The latter condition ensures that the shot noise dominates
over the thermal Nyquist-Johnson noise. \cite{blanterPR00} When
$|eV| < k_{\rm B}T$, the contribution to the noise due to thermal
fluctuations is significant, and in the strictly linear response
regime, $V\rightarrow 0$, the corresponding Fano factor diverges
due to nonzero thermal noise and vanishing of the average current
density. Let us look first at the gate voltage dependence of the
Fano factor for the lowest bias voltage shown in Fig.~\ref{Fig:3},
$V=2$ mV. The Fano factor fluctuates then between values
corresponding to the sub- and super-Poissonian noise. The
super-Poissonian shot noise ($F>1$) occurs in the blockade
regions, whereas the sub-Poissonian shot noise ($F<1$) is observed
outside the blockade regions, where the sequential transport is
energetically allowed. The minima in Fig.~\ref{Fig:3} correspond
then to positions of the peaks in Fig.~\ref{Fig:1}, and {\it vice
versa} the maxima in Fig.~\ref{Fig:3} correspond to the regions
between the peaks in Fig.~\ref{Fig:1}. It is noteworthy that this
type of correlation strongly resembles that reported for a
zero-dimensional conventional (InAs) quantum dots. \cite{nauen}

When the bias voltage increases (see Fig.~\ref{Fig:3} for $V=4$
mV), the shot noise becomes sub-Poissonian, except for those gate
voltage regions, where sequential transport is still suppressed.
This takes place only for gate voltages around the centers of the
large diamonds from Fig.~\ref{Fig:2}. When the bias voltage
increases further (see the curves for $V=8$ mV and $V=12$ mV in
Fig.~\ref{Fig:3}), the system is beyond the blockade regime
independently of the gate voltage. The corresponding Fano factor
is always smaller than unity, although it fluctuates with
increasing the gate voltage. The suppression of the Fano factor
below $F=1$ in the sequential transport regime is a consequence of
the Coulomb and exchange correlations in electronic transport.
This behavior is different from that observed in the blockade
regime, where $F>1$. \cite{onac06} However, one should bear in
mind that the sequential transport is negligible in the blockade
regions and the current is dominated there by higher-order (e.g.
cotunneling) contributions. Thus, the Fano factor in the blockade
regions may be significantly modified (in particular reduced) by
the cotunneling current. This, however, is beyond the scope of
this paper, in which the considerations are limited to the
first-order tunneling processes.

\begin{figure}[t]
  \includegraphics[width=0.49\columnwidth,height=7cm]{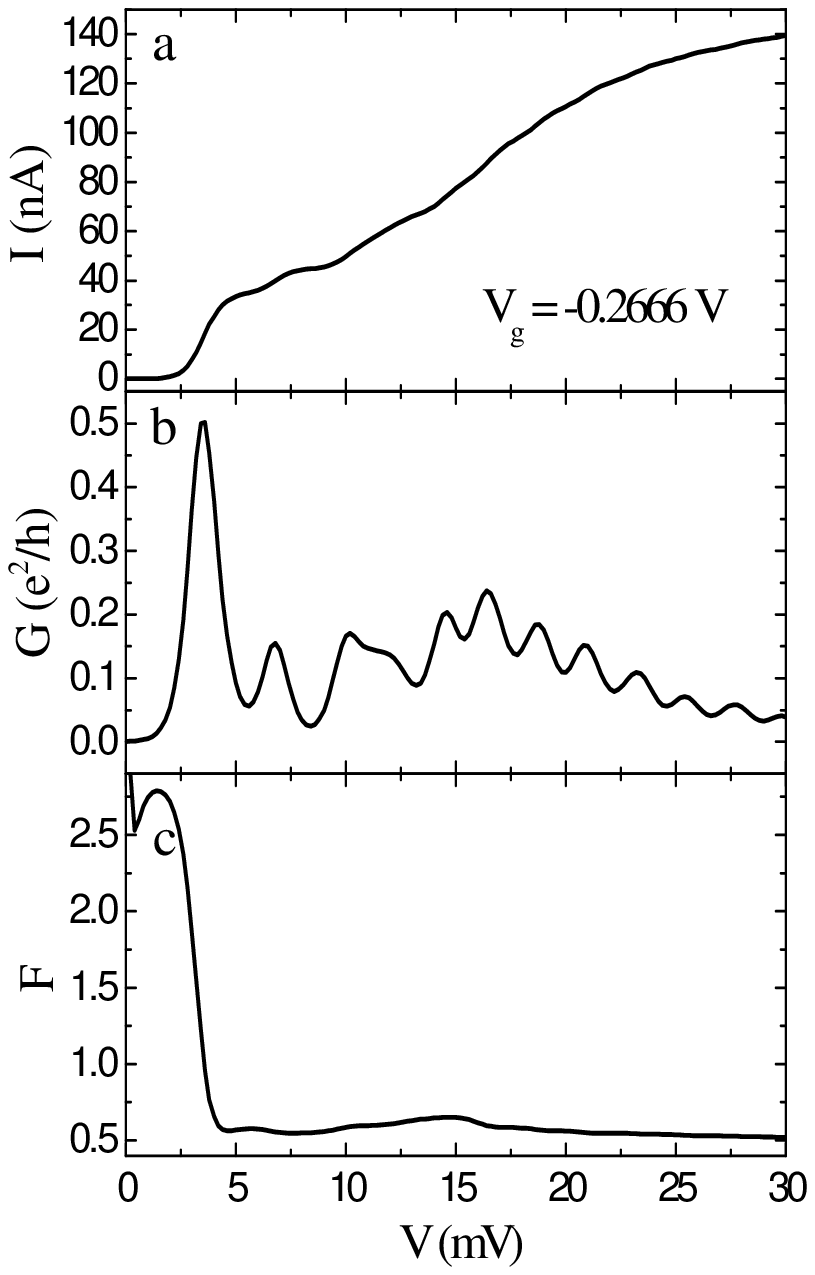}
  \includegraphics[width=0.49\columnwidth,height=7cm]{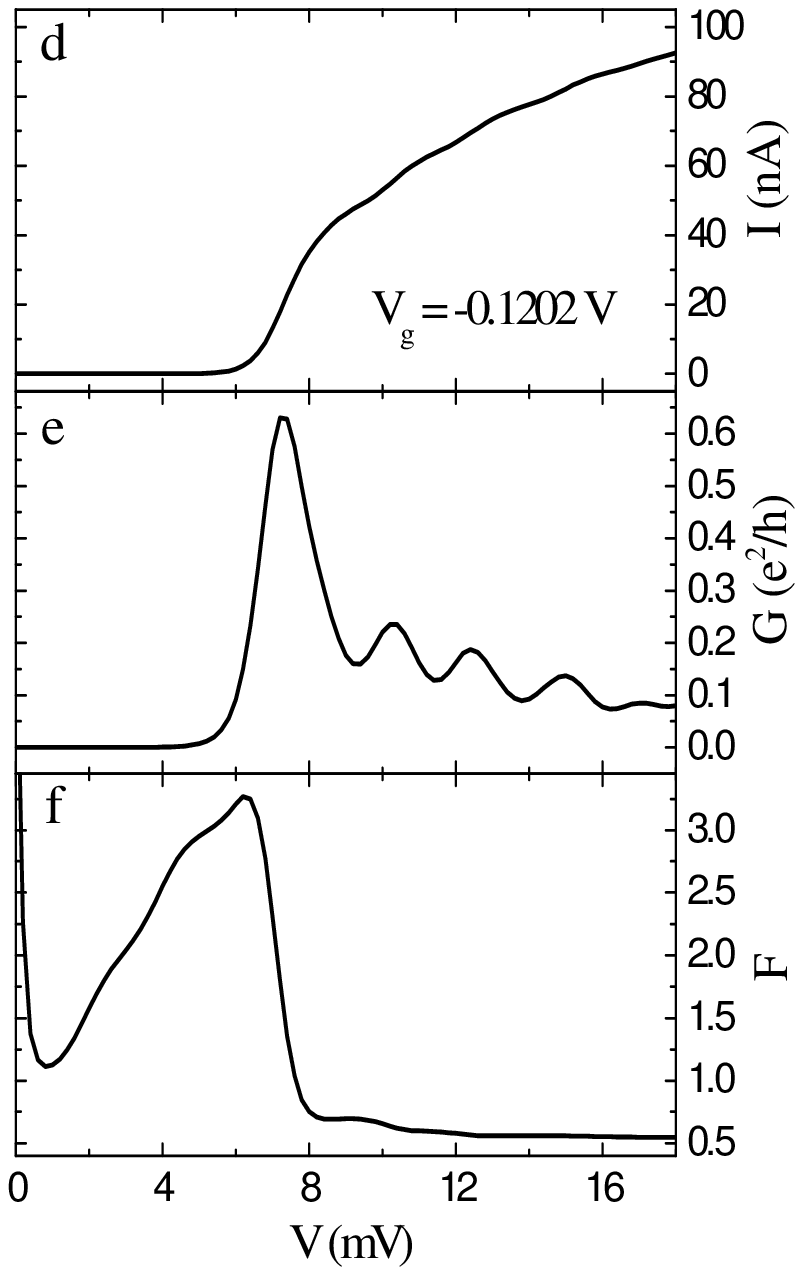}
  \caption{ \label{Fig:4}
  Bias dependence of the current (a,d), differential conductance
  (b,e) and the Fano factor (c,f) for $V_g = -0.2666$ V (left panel)
  and for $V_g = -0.1202$ V (right panel).
  The parameters are the same as in Fig.~\ref{Fig:1}. }
\end{figure}

The bias dependence of the Fano factor is explicitly shown in
Fig.~\ref{Fig:4} for two different values of the gate voltage. The
first value (left column in Fig.~\ref{Fig:4}), $V_g = -0.2666$ V,
corresponds to the center of the small diamond shown in
Fig.~\ref{Fig:2}, whereas the second value (right column in
Fig.~\ref{Fig:4}), $V_g = -0.1202$ V, corresponds to the center of
the large diamond. In order to see a clear correlation with
transport properties, the current and differential conductance are
also shown there. Firstly, the regions where the current is
suppressed are clearly visible, while for voltages above the
threshold voltage electrons tunnel one by one through the system
giving rise to the current, see Figs.~\ref{Fig:4}a and d.
Secondly, by increasing the bias voltage, one increases the number
of nanotube charge states participating in transport. This leads
to a number of peaks in the differential conductance, which can be
seen in Figs.~\ref{Fig:4}b and e. On the other hand, the
corresponding Fano factor is displayed in Figs.~\ref{Fig:4}c and
f. As already mentioned above, the Fano factor diverges in the
limit of zero bias due to thermal noise. When $|eV| \approx k_{\rm
B}T$, the shot noise becomes dominant, and the Fano factor is then
significantly reduced. However, it is still super-Poissonian in
the regions, where the current is exponentially suppressed
(blocked either by the charging energy or by the absence of
discrete levels available for tunneling in the tunneling window).
The factor $F$ becomes significantly reduced below the Schottky
value, $F<1$, for the bias voltages large enough to set the system
beyond the blockade regions. The corresponding Fano factor is then
slightly above 0.5, which clearly shows suppression of the shot
noise in comparison to the Poissonian transport processes, and
also indicates the important role of Coulomb correlations in
transport.

The super-Poissonian shot noise in the Coulomb blockade regime is
particularly pronounced close to the threshold voltage above which
the sequential transport is allowed, see Fig. \ref{Fig:4}c and f.
This enhancement can be accounted for as follows. Assume that the
voltage is slightly below the threshold one (close to the maximum
of the peak in the Fano factor). The system is then in the
blockade regime, where the current is exponentially suppressed. A
nonzero small current can flow due to thermal excitations. There
is an exponentially small probability that one electron leaves the
CNT (reducing its charge state) and then another electron tunnels
to the CNT, either to the same energy level or to a level of
higher energy. If it tunnels to the level of higher energy,
transport through this level is allowed and the electron can
easily leave the nanotube, while another one can jump to the same
level or to the level of lower energy. If it tunnels to the same
level, further tunneling processes are allowed. If it tunnels to
the low energy level, the system becomes blocked again. Similar
scenario also holds when an electron tunnels first to the CNT
(increasing its charge state) and then leaves the CNT. All this
leads to large fluctuations in the current, and consequently to
super-Poissonian shot noise. However, as we have already mentioned
above, the sequential transport is negligible in the blockade
regions and the current is dominated there by higher-order (e.g.
cotunneling) contributions (not considered here). Thus, the Fano
factor in the blockade regions may be significantly modified
(reduced) by the cotunneling current.

\begin{figure}[h]
  \includegraphics[width=0.65\columnwidth]{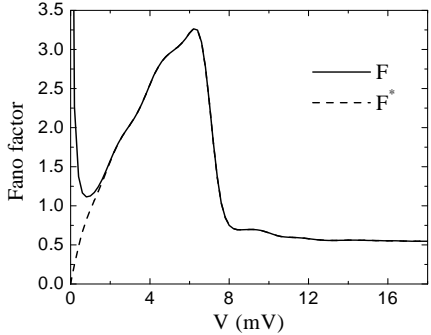}
  \caption{\label{Fig:5}
  The Fano factor, calculated using the total noise (solid line)
  and the excess noise (dashed line), as a function
  of the bias voltage for $V_g = -0.1202$ V.
  The parameters are the same as in Fig.~\ref{Fig:1}.}
\end{figure}

We also notice that experimentally one usually measures the excess
noise, $S^\star = S(V) - S(V=0)$. The corresponding Fano factor is
then defined as, $F^\star = S^\star/(2e|I|)$. In Fig.~\ref{Fig:5}
we show the bias voltage dependence of the Fano factor $F$ for
$V_g = -0.1202$ V, calculated using the total shot noise, $S$, and
$F^\star$ calculated using the excess noise, $S^\star$. One can
see that in the sequential tunneling regime the two definitions
yield equal Fano factors, $F \approx F^\star$. This is due to the
fact that the noise due to sequential tunneling in the Coulomb
blockade regime is negligible as compared to the noise above the
threshold voltage, $S(V=0) \ll S(V)$. Interestingly, the
super-Poissonian behavior observed in the Coulomb blockade is also
present irrespective of the definition of the Fano factor. The
only difference between $F$ and $F^\star$ can be seen in the low
bias voltage regime where $F \rightarrow \infty$ and $F^\star
\rightarrow 0$, as $V \rightarrow 0$, see Fig.~\ref{Fig:5}.
Therefore, in this paper we only show and discuss the Fano factor
calculated from the total shot noise, $S$, while we notice that
the behavior of $F$ exactly reflects the behavior of $F^\star$,
except for $V \rightarrow 0$.

\section{Carbon nanotubes coupled to ferromagnetic leads}

Carbon nanotubes connected to ferromagnetic leads are of
particular importance for modern magnetoelectronics.
\cite{cottet06,schonenberger06} It is worth noting that the
magneto-electronics based on all-metal multilayers has been making
impressive progress over the last two decades. Owing to the
discovery of the giant and tunnel magnetoresistance effects, the
multilayer-based devices are now wide-spread in the modern
electronics (nonvolatile memory elements, angular meters,
read/write heads, etc.). Intensive studies have also been carried
out on all-semiconductor magnetic systems, since the discovery of
ferromagnetic semiconductors. \cite{ohno} The hybrid systems
consisting of ferromagnetic metals or semiconductors and
molecules\cite{sanvito06,naber} open a new area in modern
magnetoelectronics, and therefore draw attention of physicists,
engineers and technologists. A very promising example of magnetic
metal/molecule junctions are ferromagnetically contacted CNTs,
studied in this section.

\begin{figure}[t]
  \includegraphics[width=0.85 \columnwidth]{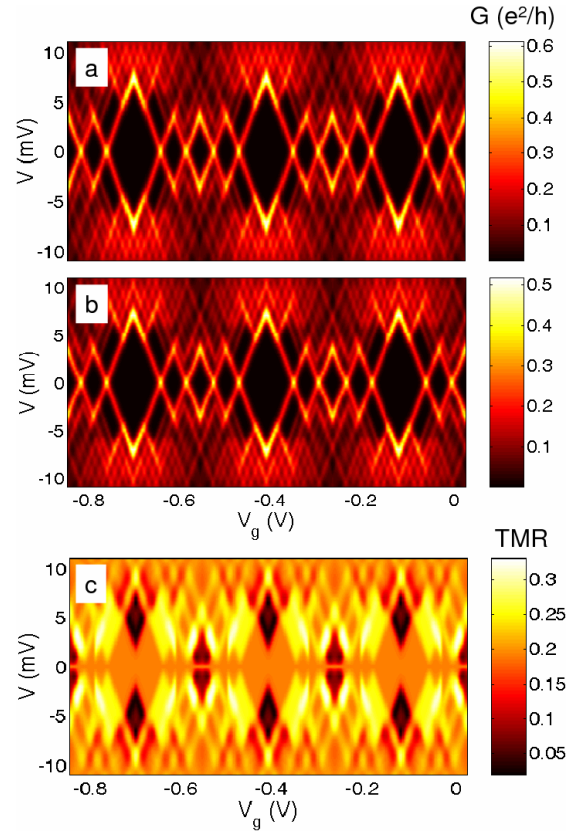}
  \caption{\label{Fig:6} (color online)
  Differential conductance in the parallel (a)
  and antiparallel (b) magnetic configurations
  and tunnel magnetoresistance (c)
  as a function of bias and gate voltages.
  The parameters are the same as in Fig.~\ref{Fig:1}
  except that $p_{\rm L} = p_{\rm R} = 0.4$.}
\end{figure}

We consider thus electronic transport in the situation when the
CNT is weakly coupled to two ferromagnetic leads. The charge
current depends then on magnetic configuration of the system, i.e.
on the relative alignment of the leads' magnetic moments.
Differential conductance in the parallel and antiparallel magnetic
configurations as a function of the bias and gate voltages is
shown in Fig.~\ref{Fig:6}a and b. As one can note from the
corresponding scales, the conductance is smaller in the
antiparallel configuration. This is a general behavior,
characteristic of the so-called normal spin valves (normal spin
valve effect), and results from the spin asymmetry in tunneling
processes. Apart from this difference, the general behavior of
differential conductance with transport and gate voltages in both
configurations is qualitatively similar, with characteristic
diamonds corresponding to blockade regions and four-fold
periodicity in the linear response regime (similar to those found
in CNTs contacted to nonmagnetic leads, see Fig.~\ref{Fig:2}).

\begin{figure}[t]
  \includegraphics[width=0.65\columnwidth]{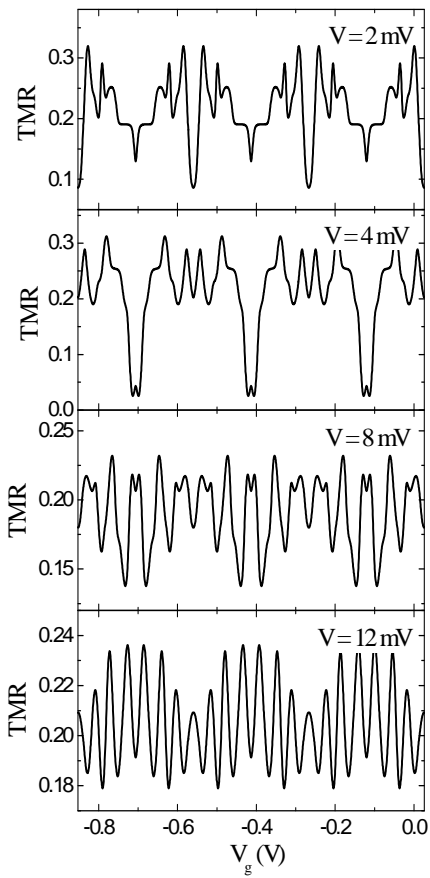}
  \caption{\label{Fig:7}
  Tunnel magnetoresistance as a function of gate
  voltage for several values of the bias voltage.
  The parameters are the same as in Fig.~\ref{Fig:1}
  with $p_{\rm L} = p_{\rm R} = 0.4$. Tunnel magnetoresistance in the linear response
  is given by ${\rm TMR} = p_{\rm L}p_{\rm R}/(1-p_{\rm L}p_{\rm R})$.}
\end{figure}

A quantity which takes into account the difference between
Fig.~\ref{Fig:6}a and b, and which is of practical importance, is
the so-called tunnel magnetoresistance (TMR). The TMR phenomenon
consists in a change of the tunneling current when magnetic
configuration varies from antiparallel to parallel alignment, and
is quantitatively defined as \cite{julliere75,barnas98}
\begin{equation}
   {\rm TMR} = \frac{I_{\rm P} - I_{\rm AP}}{I_{\rm AP}} \,,
\end{equation}
where $I_{\rm P}$ ($I_{\rm AP}$) is the current flowing through
the biased system in the parallel (antiparallel) magnetic
configuration. The dependence of TMR on the transport and gate
voltages is shown in Fig.~\ref{Fig:6}c. From this plot one can
conclude that TMR in the strictly linear response regime
($V\rightarrow 0$) is constant, i.e. independent of the gate
voltage. More precisely, it is equal to a half of the Julliere
value \cite{julliere75} for the corresponding planar junction,
i.e. ${\rm TMR} = p_{\rm L}p_{\rm R}/(1-p_{\rm L}p_{\rm R})$. This
behavior is rather general for electronic transport through
quantum dots when only first-order transport processes are taken
into account, and holds no longer when higher-order processes are
included -- especially in the blockade regions.
\cite{weymannPRB05} We note again, that including higher-order
contributions may significantly change the TMR in the blockade
regions, i.e. in the areas of Fig.~\ref{Fig:6}c corresponding to
the black diamonds in Fig.~\ref{Fig:6}a and b. Generally, one may
expect some enhancement of TMR in these regions, similarly to the
enhancement of TMR due to cotunneling processes, observed in other
magnetic tunnel junctions including also granular structures.

As follows from Fig.~\ref{Fig:6}c, TMR is no longer constant in
the nonlinear transport regime. This is shown explicitly in
Fig.~\ref{Fig:7}, where the gate voltage dependence of TMR is
presented for several values of the bias voltage. The TMR is
always positive for the assumed values of the parameters, and
oscillates with the gate voltage (contrary to the linear response
regime, where TMR is constant). The amplitude of the oscillations,
however, decreases with increasing the bias voltage.

\begin{figure}[t]
  \includegraphics[width=0.65\columnwidth]{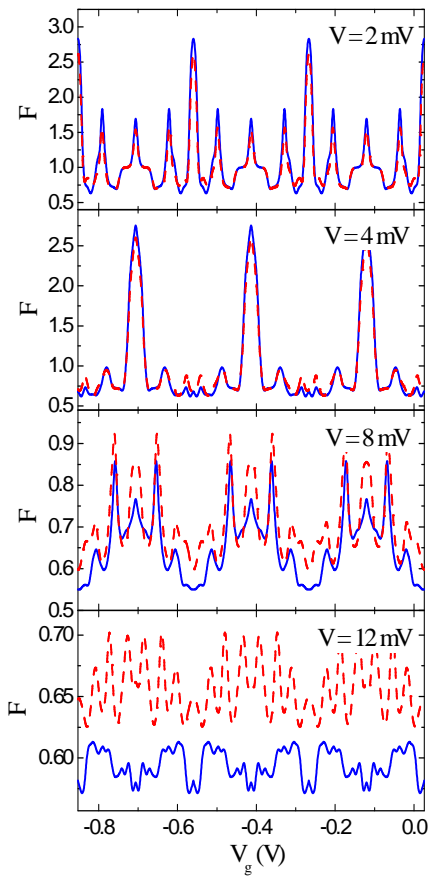}
  \caption{\label{Fig:8} (color online)
  The Fano factor in the parallel (solid)
  and antiparallel (dashed) magnetic configurations
  as a function of gate
  voltage for several values of the bias voltage.
  The parameters are the same as in Fig.~\ref{Fig:1}
  with $p_{\rm L} = p_{\rm R} = 0.4$.}
\end{figure}

The gate voltage dependence of the Fano factor in the parallel and
antiparallel magnetic configurations is shown in Fig.~\ref{Fig:8}
for a few values of the bias voltage. The general features of the
Fano factor with increasing bias and gate voltages are similar to
those observed in the case of CNTs contacted to nonmagnetic
electrodes. As before, the noise is super-Poissonian in the
blockade regions, and sub-Poissonian outside the blockade regions.
However, the Fano factor $F$ in the antiparallel configuration is
different from that in the parallel one, and may be either larger
or smaller than the latter. This difference is particularly
pronounced for larger values of the bias voltage, where the system
is out of the blockade regions. We note that for the value of the
spin polarization of the leads assumed in calculations, the Fano
factor in the antiparallel configuration, $F_{\rm AP}$, is larger
than the Fano factor in the parallel configuration, $F_{\rm P}$.
The ratio of $F_{\rm P}/F_{\rm AP}$ however depends on the spin
polarization of the leads, and for half-metallic electrodes
$F_{\rm P}$ becomes super-Poissonian, while $F_{\rm AP}$
approaches unity. This behavior is similar to that observed in
quantum dots. \cite{bulkaPRB00,cottetPRB04,weymannJPCM07}

\begin{figure}[t]
  \includegraphics[width=0.49\columnwidth,height=9cm]{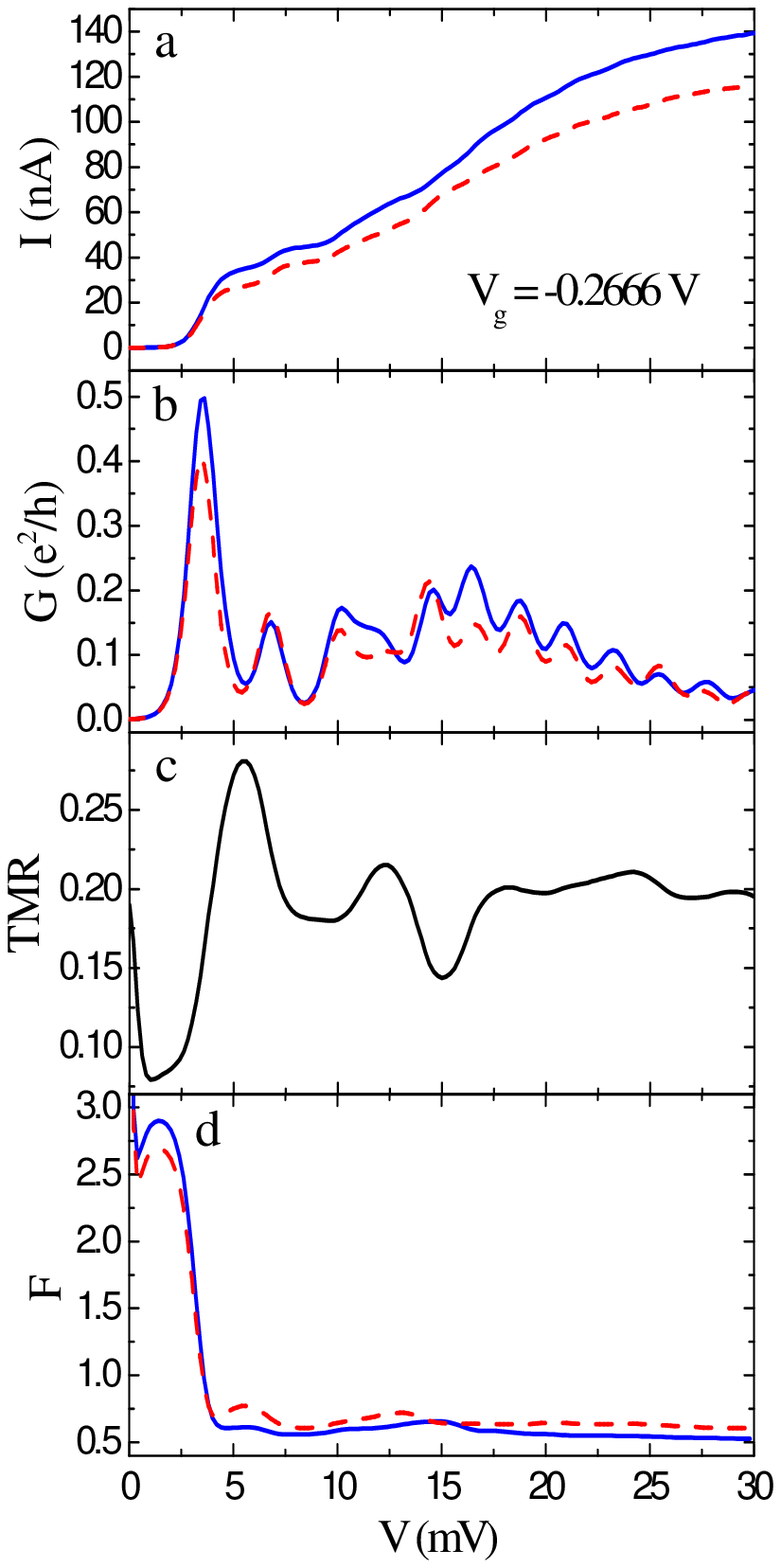}
  \includegraphics[width=0.49\columnwidth,height=9cm]{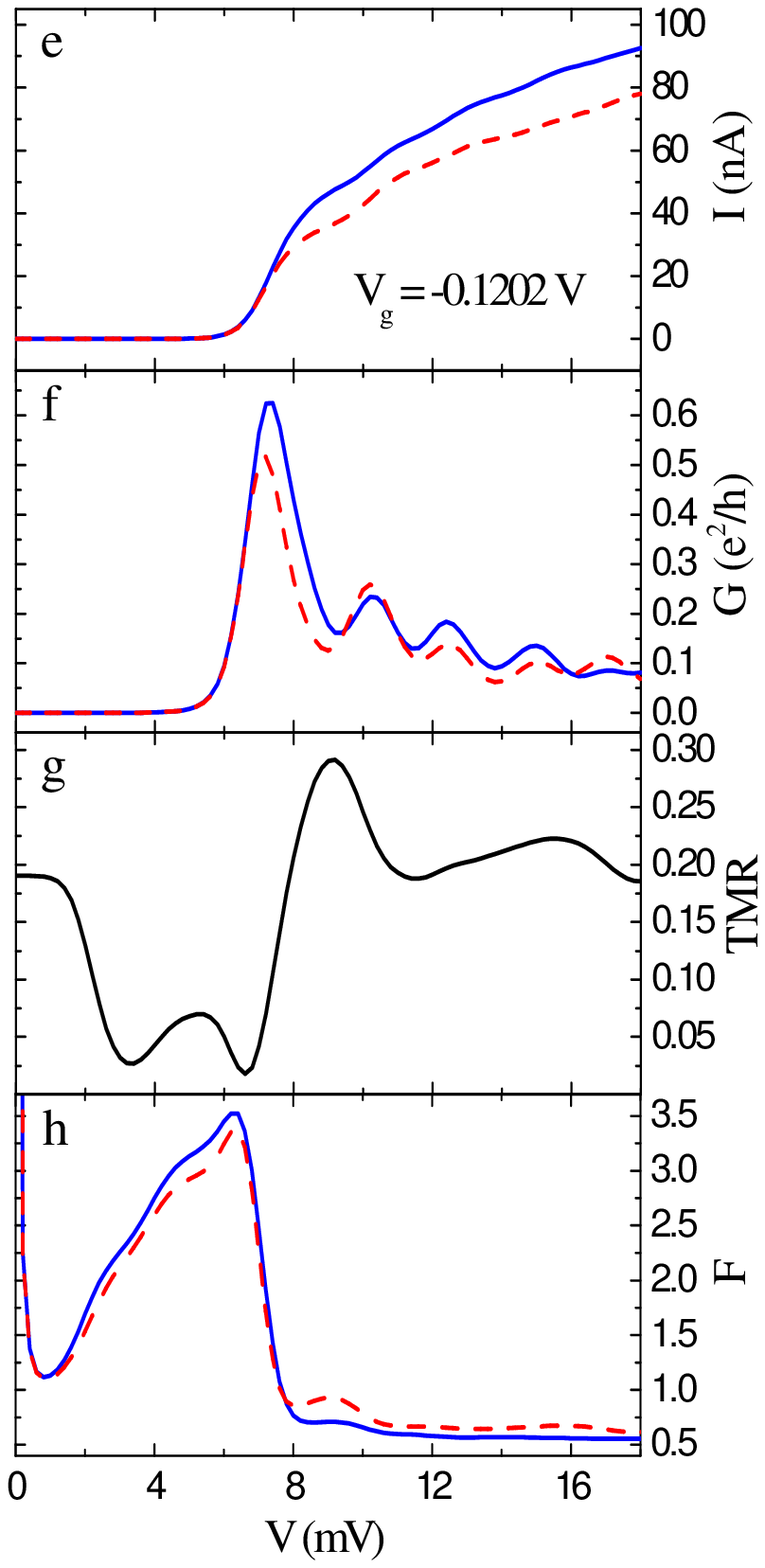}
  \caption{\label{Fig:9}
  (color online) Bias dependence of the current (a,e), differential conductance (b,f),
  TMR (c,g), and Fano factor (d,h) for $V_g = -0.2666$ V (left column)
  and for $V_g = -0.1202$ V (right column),
  and $p_{\rm L} = p_{\rm R} = 0.4$. The solid (dashed) line in
  (a,e), (b,f) and (d,h) corresponds to the parallel (antiparallel)
  magnetic configuration.
  The other parameters are the same as in Fig.~\ref{Fig:1}.}
\end{figure}

The bias voltage dependence of the Fano factor is shown explicitly
in Fig.~\ref{Fig:9} for two different values of the gate voltage
chosen so that the first (second) value corresponds to the middle
of the small (large) diamond in Fig.~\ref{Fig:6}. Figure
\ref{Fig:9} clearly shows that the shot noise is sub-Poissonian,
except for the blockade regions, where the noise is enhanced above
the Schottky value. This behavior is similar to that observed in
CNTs contacted to nonmagnetic leads, and its physical origin is
the same as it does not result from magnetism of the electrodes.
The effect is only quantitatively modified by ferromagnetism of
the electrodes -- the magnitude of the super-Poissonian shot noise
is different in parallel and antiparallel configurations, as shown
in Figs.~\ref{Fig:8} and \ref{Fig:9}. In addition, in
Figs.~\ref{Fig:9}a and e we also show the current flowing through
the system in both magnetic configurations. The difference in
currents, resulting from spin asymmetry in tunneling processes in
the antiparallel configuration, leads to nonzero TMR displayed in
Figs.~\ref{Fig:9}c and g. On the other hand, the bias dependence
of the differential conductance displays characteristic peaks,
indicating the different charge states of the nanotube being
active in transport, see Figs.~\ref{Fig:9}b and f.

\section{Summary and conclusions}

Using the real-time diagrammatic approach we have analyzed
transport through CNTs contacted to nonmagnetic and also
ferromagnetic electrodes in the weak coupling regime, where the
effects due to charging of CNT with individual electrons are
clearly seen in transport characteristics. Numerical results on
the conductance in linear response regime reproduce quite well the
experimentally observed four-peak patterns. Generally, conductance
in both the linear and nonlinear transport regimes reveals
features of the discrete electronic structure of the CNT-based
quantum dots. When the CNT is contacted to ferromagnetic leads,
the difference in conductance in the parallel and antiparallel
configurations leads to TMR, which also reveals characteristic
features of the discrete charging with single electrons and
discrete intrinsic electronic structure of the CNTs.

We have also found that in the sequential tunneling regime the
shot noise of the single wall metallic carbon nanotubes is
generally suppressed below the Schottky value, except for the
blockade regions, where it is larger than the Schottky value. For
voltages above the threshold for sequential tunneling the
corresponding Fano factor has been found to be slightly above 0.5,
while in the blockade regions it is larger than 1. However, we
expect that the shot noise in the blockade regions will be
strongly modified by higher-order contributions to the current. It
has also been shown that for the assumed parameters the Fano
factor in the antiparallel configuration is typically larger than
the Fano factor in the parallel configuration.


\begin{acknowledgments}
This work was supported by: the EU grant CARDEQ under contract
IST-021285-2, and the Polish Ministry of Science and Higher
Education as a research project in years 2006-2009. I.W. also
acknowledges support from the Foundation for Polish Science.
\end{acknowledgments}


\end{document}